\documentclass[12pt, a4paper]{article}
\pdfoutput=1
\usepackage[latin1]{inputenc}
\usepackage{amsmath}
\usepackage{amsfonts}
\usepackage{amssymb}
\usepackage{latexsym}
\usepackage{mathrsfs}
\usepackage{graphicx}
\usepackage{color}
\usepackage{twistor}
\usepackage[all]{xy}
\usepackage[colorlinks=true, linkcolor=black, citecolor=black, urlcolor=black, linktocpage=true]{hyperref}

\renewcommand{\d}{\mathrm{d}}

\numberwithin{equation}{section}

\begin{document}

\title{\Large Einstein supergravity amplitudes from twistor-string theory}

\author{\normalsize Tim Adamo \& Lionel Mason \\ \small \textit{The Mathematical Institute} \\ \small \textit{University of Oxford} \\ \small \textit{24-29 St Giles'} \\ \small \textit{Oxford, OX1 3LB, U.K.}}

\date{}

\maketitle

\abstract{This paper gives a twistor-string formulation for tree amplitudes of Einstein (super-)gravities for $\cN=0$ and $4$.  Formulae are given with and without cosmological constant and with various possibilities for the gauging.  The formulae are justified by use of Maldacena's observation that conformal gravity tree amplitudes with Einstein wave functions and non-zero cosmological constant will correctly give the Einstein tree amplitudes.  This justifies the construction of Einstein gravity amplitudes at $\cN=0$ from twistor-string theory and is extended to $\cN=4$ by requiring the standard relation between the MHV-degree and the degree of the rational curve for Yang-Mills; this  systematically excludes the spurious conformal supergravity gravity contributions.   For comparison, BCFW recursion is used to obtain twistor-string-like formulae at degree zero and one (anti-MHV and MHV) for amplitudes with $\cN=8$ supersymmetry with and without cosmological constant.}

\pagebreak

\tableofcontents


\section{Introduction}
\label{intro}

Perhaps one of the most striking formulae to emerged from Witten's twistor-string theory \cite{Witten:2003nn} was the formula for the tree-level S-matrix of $\cN=4$ super Yang-Mills as an integral over rational curves in twistor space \cite{Roiban:2004yf}.  The formula is 
\be{gauge-tree}
\cA(1,\ldots,n)=\int_{\CM_{d,n}} \d\mu_d \prod_{i=1}^n\frac{a_i(Z(\sigma_i))\d\sigma_i}{\sigma_i-\sigma_{i-1}}
\ee
where $a_i$ are the twistor space wave functions for the scattered particles, the integral is over the space $\CM_{d,n}$ of maps of a rational curve to twistor space of degree $d$ with $n$ marked points, $\sigma$ is a coordinate on the rational curve and  $Z(\sigma)$ the map from the rational curve to twistor space.
Although the detailed definitions are left until later, it is nevertheless clear that this remains the most succinct formula for gauge theory tree amplitudes that manifests the most symmetries.

Twistor-string theory also contains $\cN=4$ conformal supergravity
\cite{Berkovits:2004jj} and analogous, albeit more complicated,
formulae can be obtained for conformal supergravity amplitudes
\cite{Ahn:2005es,Dolan:2008gc}.  These formulae have
potential extensions to loop amplitudes although the anomalies are
still not well understood, see \cite{Berkovits:2004hg,
 Mason:2007zv, Dolan:2007vv}.   The presence of conformal supergravity
has been the most significant obstruction to applying twistor-string
theory to gauge theory amplitudes as, naively at least, there would be
no mechanism for preventing conformal supergravity modes from running
around the loops and corrupting the answers.   On the other hand, this presents an opportunity, as conventional gravity sits inside conformal gravity
at the level of the field equations.  Indeed this was part of the
motivation for the string theories for Einstein
supergravities proposed in \cite{AbouZeid:2006wu}, where a gauging was
introduced to restrict the spectrum of conformal supergravity to that
of Einstein supergravity, the hope being that the twistor-string
theory would then correctly produce the amplitudes.  However, these
twistor-string theories 
turned out to be untenable descriptions of Einstein gravity as the gauging effectively eliminated the instanton sectors, reducing the theories to chiral interactions \cite{Nair:2007md, Brodel:2009ep}. 

Recently, Maldacena \cite{Maldacena:2011mk} has shown that if one
considers gravity with a non-zero positive cosmological constant
$\Lambda$ embedded in conformal gravity, then the conformal gravity
calculation of the tree-level amplitudes will give the correct result
for Einstein gravity simply using Einstein gravity scattering states.
The argument, based on earlier work of Anderson \cite{Anderson:2001},
relies on showing that the conformal gravity action evaluated on a
solution to the Einstein equations yields a fixed multiple of the
regularized Einstein action (together with the Euler characteristic).  
With de Sitter asymptotics the Einstein action is necessarily divergent, being proportional to $\Lambda$ times the volume.   However, it is now well understood how to regularise the volume (c.f., \cite{Balasubramanian:1999re, Skenderis:2002wp, Miskovic:2009bm}) and one obtains, assuming $R_{ab}=\Lambda g_{ab}$, using the Chern-Gauss-Bonnet formula 
\be{conf-grav}
\int \d^{4}x\sqrt{g}\;|W|^2=8\pi^2\chi -\frac23\Lambda^2V \; .
\ee
Here, $|W|^{2}$ is the contracted square of the Weyl curvature, $V$ is the regularized volume and $\chi$ the Euler characteristic.
Thus, up to an irrelevant numerical factor and topological term, one obtains   
$\Lambda$ multiplied by the Einstein action.  The tree-level S-matrix is essentially the action evaluated on a perturbative solution to the equations, so we conclude that, up to this multiple of $\Lambda$, the conformal gravity calculation of amplitudes restricted to Einstein gravity in and out states will give $\Lambda $ times the Einstein amplitudes.
This correspondence will degenerate as $\Lambda$ is taken
to zero and this perhaps goes some way to explain the difficulties found
by \cite{Nair:2007md,Brodel:2009ep} with the Einstein twistor-string
theories proposed in \cite{AbouZeid:2006wu} but raises the hope that
something can be done when $\Lambda \neq 0$.  

In this article we will make the (widely believed) assumption that twistor-string theory correctly gives rise to conformal supergravity amplitudes at tree-level. Maldacena's argument will then allow us  to deduce such formulae for gravitational tree
amplitudes for $\Lambda \neq 0$.  This limit $\Lambda\rightarrow 0$ is, at least in abstract, straightforward because, by construction, an $n$-point amplitude  will be a polynomial of degree $n$ in $\Lambda$ that vanishes at $\Lambda=0$, so we can divide by $\Lambda$ to obtain the correct Einstein amplitudes.
After a brief review of twistor-string theory for conformal supergravity amplitudes, we explain how Einstein scattering states can be inserted
into the twistor-string formulae.  We examine some elementary amplitudes, at degrees zero and one,  discuss their extension  to $\cN=4$
supergravity\footnote{It is well known that de Sitter supergravity is problematic \cite{Witten:2001kn} in the sense that the supersymmetry cannot remain unbroken if represented unitarily.  This will make little difference to our perturbative calculations as all formulae are polynomial in $\Lambda$ and the sign can be taken either way.}  and the limit
$\Lambda\rightarrow 0$.  We show that the embedding of the Einstein states into the conformal supergravity degree zero amplitudes corresponds to a reduction of the self-dual twistor action of Berkovits and Witten \cite{Berkovits:2004jj} to the self-dual Einstein twistor actions of \cite{Mason:2007ct} multiplied by $\Lambda$.   We show that the some amplitudes for conformal supergravity when restricted to Einstein  supergravity wave-functions are inconsistent with Einstein supergravity, but that these can be systematically eliminated by requiring the standard relationship between the degree $d$ of the space of rational curves on which the amplitude is computed, and the MHV degree $k$ of the amplitude, $d=k+1$.   We also show that BCFW recursion at degree zero and one give rise to twistor-string-like formulae at degree zero and one (similar twistor-string-like formulae were already found for MHV amplitudes with $\Lambda=0$ in \cite{Nair:2005iv,Mason:2008jy}).  Some comments about the validity of BCFW recursion with $\Lambda\neq 0$ are included in appendix \ref{BCFWapp}.  
Thus, twistor-string tree formulae exist for Einstein supergravities with $\cN=0,4$ and at low degree for $\cN=8$.  This provides the best evidence so far for the existence of an Einstein twistor-string theory.

For the most part we use general wave functions for the scattering represented in twistor space.   In the BCFW section we use wave functions supported at a point in twistor space.   These are particularly advantageous in this context because, although gravity breaks conformal invariance, it does so in a particularly mild way: linearized spin two fields are conformally invariant and can be mapped onto linear gravity perturbations with positive, negative or vanishing cosmological constant according to taste \cite{Mason:1987}.  Twistor theory makes this conformal symmetry breaking manifest with the choice of an {\em infinity twistor} that determines the cosmological constant. This underlying conformal invariance means that we also obtain attractive formulae for the three-particle amplitudes using momentum eigenstates that are not adapted to the de Sitter group, but to a 4-dimensional translation subgroup of the conformal group appropriate to the $\Lambda\rightarrow 0$ limit of the infinity twistor.  Such three point amplitudes in de Sitter space have been studied in \cite{Maldacena:2011nz} with a conjectured extension to higher numbers of points in \cite{Raju:2012zr} made explicit at four points in \cite{Raju:2012zs}.  However those momentum eigenstates are tied to a different choice of translation subgroup of the conformal group, and so are difficult to compare except in their $\Lambda=0$ limit; even there, there are additional (singular) factors in their formulae, whereas ours have smooth limits at $\Lambda=0$.  Nevertheless, theirs, being tied to the geometry of infinity in AdS, are useful for studying gravity amplitudes in the AdS/CFT correspondence.

An alternative approach to these ideas arises from the twistor action for conformal supergravity \cite{Mason:2005zm,Mason:2007zv}.  This can be thought of as giving the string field theory for twistor-string theory.  Again the Maldacena argument shows that, up to a factor of $\Lambda$,  these will give the correct amplitudes for Einstein gravity at tree-level when restricted to Einstein wave functions.  
A more complete exposition of these findings which also treats these issues from the perspective of the space-time and twistor actions extending and consolidating \cite{Mason:2008jy} for gravity with a cosmological constant will appear in \cite{Adamo:2012}.


\section{Twistor-strings for conformal supergravity}

Nonprojective twistor space is $\T=\C^{4|4}$ and projective twistor
space is $\PT=\T/\{Z\sim \e^\alpha Z\}$, $\alpha\in\C$.  A twistor will be
represented as $Z^I\in\T$, $Z^I=(Z^\alpha,\chi^a)$, $\alpha=0,\dots 3$,
$a=1,\ldots,4$ with $Z^\alpha$ bosonic and $\chi^a$ fermionic and the
bosonic part $Z^\alpha=(\lambda_{A}, \mu^{A'})$, $A=0,1$, $A'=0'1'$.  A
point $(x,\theta)=(x^{AA'},\theta^{Aa})$ in chiral super Minkowski
space-time $\M$ corresponds to the $\CP^1$ (complex line) $X\subset
\PT$ via the incidence relation \be{incidence}
\mu^{A'}=ix^{AA'}\lambda_{A}\, , \quad \chi^a=\theta^{Aa}\lambda_{A}\, .
\ee with $\lambda_{A}$ homogeneous coordinates along $X$.

We use a closed string version\footnote{This is essentially equivalent to a heterotic half-twisted $(0,2)$-model of twistor-string theory \cite{Mason:2007zv}.    Such heterotic models are equivalent to a gauged linear $\beta$-$\gamma$ system using the ideas of \cite{Witten:2005px}.  This is in practice also equivalent to Witten's B-model in which the role of the D1-instantons is taken over by the fundamental string.} of  the Berkovits model\footnote{The standard Berkovits open twistor-string would be just as good for most of the considerations in this paper; although it is tied into split signature and gets some signs wrong, subtleties over choices of contours are avoided.}  \cite{Berkovits:2004hg} with a Euclidean worldsheet $\Sigma$.  This is the most natural formulation for making contact with arbitrary signatures and the standard cohomological descriptions of wave functions on twistor space, whilst exploiting the relative simplicity of the Berkovits model.  The fields are
$$
Z:\Sigma\rightarrow \T\, , \quad Y\in \Omega^{1,0}(\Sigma)\otimes T^*\T\, , \quad \mbox{ and } \quad a\in \Omega^{0,1}(\Sigma)\, .
$$ 
The action is
\be{string-action}
S[Z,Y,a]=\int_\Sigma Y_{I}\dbar Z^{I} + a Z^{I}\; Y_{I} +S_{C}\, ,
\ee
where $S_{C}$ is the worldsheet current action.  This action has the gauge freedom
$$
(Z,Y,a)\rightarrow (\e^\alpha Z,\e^{-\alpha}Y, a-\dbar \alpha)\, .
$$
The gauging reduces the string theory to one in $\PT$ and the formalism allows one to use homogeneous coordinates on $\PT$. 

The amplitudes are computed as path integrals of CFT correlators of
vertex operators on $\Sigma$.  For gravity, the vertex operators
correspond to deformations of the complex structure together with
deformations of the $B$-field  (a Hermitian $(1,1)$-form determining
the metric in the heterotic framework, although in the half-twisted
framework, reduction to $\bar Q$-cohomology means that only the
cohomology class in $H^1(\Omega^{1,0})$ is relevant).  These are given
by $\dbar$-closed $(0,1)$-forms $F$ on the bosonic part of twistor space $\PT$ with values in $ T\oplus T^* \PT$, which can be represented on $\T$ by $\bar\p$-closed $(0,1)$-forms $F:=(f^I,g_I)$ of homogeneity $(1,-1)$ satisfying $\p_I f^I=0=Z^Ig_I$, defined modulo gauge transformations $(\alpha Z^I,\p_I\beta)$.  These conditions imply that $(f^I\p_I,g_I\d Z^I) $ represents a section of $T\oplus T^*\PT$.     The corresponding vertex operators take the form
$$
V_F:= V_f+V_g:=\int_\Sigma f(Z)^IY_I + g(Z)_I\d Z^I \, .
$$
These can, in the usual way, be thought of as perturbations of the
action \eqref{string-action}.  In this context $V_f$ corresponds to a
deformation of the complex structure of the twistor space and $V_g$ to
a deformation of the Hermitian (strong Kahler  with torsion) structure \cite{Mason:2007zv}.

Conformal supergravity tree-level amplitudes arise when $\Sigma=\CP^1$ and are given by the path integral perturbed by the vertex operators.  This can be expanded in $(f,g)$.  The path-integral over the non-zero modes leads to the CFT correlation function of the polynomials in the vertex operators being taken.   The remaining integral over the zero-modes  gives the finite-dimensional path integral over the worldsheet instantons
\be{amplitudes}
\cM(1,\ldots,n)=\sum_{d=0}^\infty \int_{\CM_{d,n}} \d\mu_d \la V_{F_1} \ldots V_{F_n}\ra _d \, ,
\ee
over the space  $\CM_{d,n}$ of maps $Z:\CP^1\rightarrow\PT$ of degree-$d$ and $n$ marked points.\footnote{The rules for taking the correlators are different at different degrees, hence the subscript $d$ on the correlator.}  See \cite{Nair:2007md} for further explanation.   To be more concrete, we can represent the maps by
\be{measure}
Z(\sigma)=\sum_{r=0}^d \sigma_0^r\sigma_1^{d-r} U_r\, , \quad \d \mu_d =\frac{1}{\mathrm{vol}\GL(2,\C)} \prod_{r=0}^d \d^{4|4}U_r \;,
\ee
where $\sigma_{A}$ are homogeneous coordinates on $\CP^1$ and $U_r\in \T$ provide a set of coordinates on $\CM_{d,0}$ with redundancy $\GL(2,\C)$ acting on the $\sigma$ and hence the $U_r$.  The vertex operator $V_{F_i}=V_F(Z(\sigma_i))$ is inserted at the $i$th marked point $\sigma_i\in\Sigma$, and the correlator  naturally introduces a $(1,0)$-form at each marked point either from the $Y_{I}$ or the $\d Z^I$, whereas the `wave-functions' $(f^I,g_I)$ naturally restrict to give a $(0,1)$-form at each marked point.

The correlators are computed by performing Wick contractions of all the $Y$s with $Z$s to give the propagator
\be{OPE}
\la Y(\sigma)_IZ^J(\sigma')\ra= \left(\frac{\xi\cdot\sigma'}{\xi\cdot\sigma}\right)^{d+1}\frac{\delta^J_I \D \sigma}{\sigma\cdot\sigma'} \, ,\quad \D\sigma =\sigma\cdot \d\sigma\, , \quad \sigma\cdot\sigma'= \sigma_0\sigma_{1}'-\sigma_{1}\sigma_{0}'\, .
\ee
When $Y$ acts on a function of $Z$ at degree $d$, it then differentiates before applying the contraction; $Y$ acting on the vacuum gives zero so that all available $Y$s must be contracted, but this contraction can occur with any available $Z$.  The $\xi$ is an arbitrary point on the Riemann sphere and reflects the ambiguity in inverting the $\dbar$-operator on functions of weight $d$ on $\Sigma$.  The overall formula should end up being independent of the choice of $\xi$.

Unlike the Yang-Mills case, the degree $d$ of the map is not directly related to the MHV degree of the amplitude (which essentially counts the number of negative helicity gravitons minus 2).  The MHV degree of an amplitude  counts  the number of insertions of $V_g$ minus 2 (so the MHV amplitude has two $V_g$s).   Conformal supergravity amplitudes have been calculated from this formula in \cite{Ahn:2005es,Dolan:2008gc}.


\section{Reduction to Einstein Gravity}
Even for Einstein gravity without supersymmetry, we will need to use
the supergeometry of $\cN=4$ supertwistor space.\footnote{At
  tree-level we can pull out the pure gravity parts of the amplitude,
  but their construction still relies on the fermionic integration
  built into the twistor-string formulae.}  
We first give the restriction required of the vertex operators for $\cN=4$ supersymmetry and then for $\cN=0$.

To reduce to the Einstein case we must break conformal invariance.
This is done by introducing skew infinity twistors $I_{IJ}$, $I^{IJ}$ with super-indices $I=(\alpha, a)$.
The bosonic parts $I_{\alpha\beta}$, $I^{\alpha\beta}$ satisfy  
$$
I^{\alpha\beta}=\frac12\varepsilon^{\alpha\beta\gamma\delta}I_{\gamma\delta}\, , \quad I^{\alpha\beta}I_{\beta\gamma}=\Lambda \delta^\alpha_\gamma\; ,
$$ 
where $\Lambda$ is the cosmological constant.  In terms of the
spinor decomposition of a twistor $Z^\alpha=(\lambda_{A}, \mu^{A'})$ we
have
\be{infinity}
I_{\alpha\beta}=\begin{pmatrix}  \varepsilon^{AB}& 0\\ 0 &
 \Lambda \varepsilon_{A'B'}\end{pmatrix}\, , \qquad
I^{\alpha\beta}=\begin{pmatrix} \Lambda\varepsilon_{AB} & 0\\ 0 & \varepsilon^{A'B'}\end{pmatrix}\:.
\ee
They have rank two when $\Lambda=0$
(i.e., the cosmological constant vanishes) and four otherwise.

The geometrical interpretation of $I_{\alpha\beta}$ is that lines in
$\PT$ on which $I_{\alpha\beta}Z^\alpha \d Z^\beta$ vanishes correspond
to points at infinity. This can be expressed more directly by writing 
a line in twistor space as a bi-twistor
$X^{\alpha\beta}=A^{[\alpha}B^{\beta]}$ where $A,B$ are a pair of
points on the line.  Normalizing against the $\Lambda=0$ infinity
twistor, we have
\be{position}
X^{\alpha\beta}=\begin{pmatrix} \varepsilon_{AB} & ix_{A}^{B'}\\ -ix^{A'}_{B} & -\frac{1}{2}x^{2}\varepsilon^{A'B'}\end{pmatrix}\, .
\ee
Then infinity $\scri$ is the surface $X^{\alpha\beta}I_{\alpha\beta}=0$ which gives $2-\Lambda x^2=0$ in these coordinates.  This is null when $\Lambda=0$, space-like for $\Lambda >0$ (the de Sitter case), and timelike for $\Lambda <0$ (the AdS case).
The metric in these coordinates is then, with our normalization,
\be{}
\d s^2=\frac{\varepsilon_{\alpha\beta\gamma\delta}\d X^{\alpha\beta}\d X^{\gamma\delta}}{(I\cdot X)^2}=
\frac{4\d x^{AA'}\d x_{AA'}}{(2-\Lambda x^2)^2}\, .
\ee
In the supersymmetric cases, the fermionic parts of $I^{IJ}$ can be non-zero and correspond to some gauging of the $R$-symmetry of the supergravity \cite{Wolf:2007tx,Mason:2007ct}.  

Geometrically $I^{IJ}$ and $I_{IJ}$ respectively define a Poisson structure $\{,\}$ of weight $-2$ and contact structure $\tau$ of weight $2$ by
$$
\{h_1,h_2\}:=I^{IJ}\p_I h_1\p_J h_2\, , \qquad \tau=I_{IJ}Z^I\d Z^J\, ,
$$ 
and we can use the Poisson structure to define Hamiltonian vector fields $X_h=I^{IJ}\p_Ih\p_J$ which will be homogeneous when $h$ has weight 2.

The Einstein vertex operators $(V_h,V_{\tilde h})$ correspond to $V_f+V_g$ subject to the restriction $(f^I,g_I)= (I^{IJ}\p_J h, \tilde{h} I_{IJ}Z^J)$ so that
\be{Einstein-V}
 V_{h}=\int_\Sigma I^{IJ}Y_I \p_J h\, , \quad V_{\tilde{h}}=\int_\Sigma \tilde{h} \wedge \tau\, .
\ee
This is perhaps most easily seen in the non-supersymmetric case by
observing that the deformation of the complex structure  used in the
nonlinear graviton construction arises from
$f^\alpha=I^{\alpha\beta}\p_\alpha h$, see
\cite{Penrose:1976jq,Ward:1980am} and these are what leads to the
self-dual Einstein deformations. 
Supersymmetric cases are discussed in \cite{Wolf:2007tx,Mason:2007ct}, including cosmological constant where this is simply extended to $f^I= I^{IJ}\p_J h$.  In these supersymmetric extensions, nontrivial fermionic components of the $I^{IJ}$ are allowed and correspond to the gauging of
(parts of) the R-symmetry of the supergravity.  We then use the
twistor transform to deduce the corresponding formula for $g_I$
\cite{Mason:1987} (see also below).  

In terms of component fields 
\begin{eqnarray*}
h &=& e_2+\chi^a \rho_{1a}+\chi^a\chi^b a_{2ab}+ \chi^3_a\psi_{-1}^a +\chi^4 \phi_{-2}\, ,\\ \tilde{h}&=& \tilde{\phi}_{-2} +\chi^a \tilde{\psi}_{-3a}+\chi^a\chi^b \tilde{a}_{-4ab}+\chi^3_a \tilde{\rho}_{-5}^a + \chi^4  \tilde{e}_{-6}\, ,
\end{eqnarray*}
where $\chi^3_a=\varepsilon_{abcd}\chi^b\chi^c\chi^d$ and $\chi^4=\varepsilon_{abcd}\chi^a\chi^b\chi^c\chi^d$ and the subscripts denote the homogeneity $k$ of the twistor function which is related to the helicity of the corresponding space-time fields by $-(k-2)/2$.
The corresponding fields on space-time are most easily obtained from the integral formula
$$
\tilde{\Phi}(x,\theta):=\int_X \tilde{h}\wedge\tau\, .
$$
Assuming that $\tau$ only has bosonic components, from the dependence of $\tilde{h}$ on the $\theta$s through $\chi^a=\theta^{Aa}\lambda_{A}$  this formula gives a superfield with just four terms
$$
\tilde{\Phi}(x,\theta)=\tilde{\phi}(x)+ \theta^{Aa}\tilde{\eta}_{Aa}+\ldots + \theta^{Aa}\theta^{Bb}\theta^{Cc}\theta^{Dd}\epsilon_{abcd } \tilde{W}_{ABCD}
$$
containing an $\cN=4$ super-multiplet starting at a solution to the wave equation going down to an anti-self-dual Weyl spinor $\tilde{W}_{ABCD}$ satisfying their standard conformally invariant field equations on de Sitter space (so that for example $(\Box+\frac R6)\tilde{\phi}=0=\nabla^{AA'}\tilde{W}_{ABCD}$).
Clearly $h$ corresponds to an identical multiplet but of the opposite chirality.  

The analagous conformal gravity multiplet is 
$$
W(x,\theta)=\int_{X} g_I\d Z^I = C(x)+ O(\theta)
$$
whose leading leading component $C(x)$ satisfies the conformally invariant equation $\Box^2 C(x)=0$ \cite{Mason:1990}.   The reduction to Einstein gravity implies the restriction of $W(x,\theta)$ to the form given above satisfying a second order rather than fourth order wave operator.  With $g_I\rd Z^I=\tilde{h}\wedge \tau$ we will have that $W(x,\theta)$ vanishes on $\scri$ (as $\tau$ vanishes on lines corresponding to points of $\scri$), and this is one way of distinguishing the linearized Einstein supergravity inside linearized conformal gravity.

In order to reduce to standard non-supersymmetric Einstein gravity we must impose 
\be{0susy}
h=e_2\, , \quad \mbox{ and }\quad \tilde{h}=\chi^4 \tilde{e}_{-6}\, .
\ee
Following Maldacena's argument \cite{Maldacena:2011mk}, with non-zero cosmological constant, conformal gravity tree-amplitudes restricted to Einstein wave functions will give rise to Einstein tree amplitudes.  
Thus, evaluated on vertex operators contructed from \eqref{Einstein-V} with \eqref{0susy}, \eqref{amplitudes} leads to the construction of Einstein gravity tree amplitudes. With this restriction, we see that there is now a correlation between degree of the maps and MHV degree, as fermionic variables only come with $\tilde{h}$: since there are $4d$ fermionic integrations in the path integral for the amplitude, there must be $d$ insertions of $\tilde{h}$.
As we will see, this no longer holds for Einstein supergravity and
indeed we will be able to construct spurious amplitudes.  However, we
will nevertheless be able to characterize those amplitudes that are
appropriate to Einstein gravity.

\subsection{Amplitudes at degree 0}
\label{deg0}

Starting with no supersymmetry (i.e., with \eqref{Einstein-V} and \eqref{0susy}) the only three-point amplitude we can obtain at degree zero must arise from one $V_g$ and two $V_f$s.  
This leads to the 3-point MHV-bar amplitude. In conformal supergravity we obtain
$$
\cM^{\mathrm{C-SUGRA}}_{\overline{\mathrm{MHV}}}(f_1,f_2,g_3)=\int_\PT g_{3I}\wedge [f_1,\wedge f_2]^I\wedge\Omega\, 
$$
where $\Omega=\D Z^{3|4}$ is the super-Calabi-Yau volume form on
$\PT$.  Because one of the $Y$s must contract with the $\d Z$ on a
degree zero map (otherwise $\d Z=0$), and the other contraction leads
to derivative terms that can be integrated by parts to obtain the above formula.

We start by calculating this with $\cN=4$ supersymmetry, but to simplify considerations we will assume that $I^{IJ}I_{JK}=\Lambda \delta^I_K$ where $\Lambda$ is the cosmological constant (and $\Lambda=0$ corresponds to the standard rank-two infinity twistors).  
When Einstein vertex operators are inserted, this reduces to the formula
\be{MHV-bar-T}
\cM^{\mathrm{SUGRA}}_{\overline{\mathrm{MHV}}}(X_{h_1},X_{h_2},\tilde{h}_3\wedge\tau)= \int_\PT  \tilde{h}_3\wedge X_{\{h_1,h_2\}} \lrcorner \tau \wedge\Omega=2\Lambda \int  \tilde{h}_3 \{h_1,h_2\} \wedge \Omega \, .
\ee
The second equality follows from the general identity 
\be{reduction}
X_{h}\lrcorner \tau=Z^{I}I_{IJ}I^{JK}\p_K h=2\Lambda h\, ,
\ee
where $I^{IJ}I_{JK}=\Lambda \delta^I_K$ allows us to use the homogeneity relation $Z^I\p_I h=2h$ and not break supersymmetry.  

Restricting to the case of $\cN=0$ supersymmetry, this can be simply
evaluated on momentum eigenstates\footnote{We remark that there are no
  4-momenta for the de Sitter group, and these
  momentum eigenstates are somewhat unnatural as far as the de Sitter
  geometry is concerned; they are singular on a finite lightcone and do not
  recognise infinity.  Nevertheless, they limit nicely onto
  the standard ones for flat space as $\Lambda\rightarrow 0$ being the momenta dual to the coordinates introduced in the previous section.} with 4-momentum $P^{AA'}=p^A\tilde p^{A'}$
\be{mom}
e_k=\int_\C\frac{\d s}{s^{k+1}} \bar\delta^2(s\lambda_{A}-p_{A})\e^{s\mu^{A'}\tilde{p}_{A'}} 
\ee
where the complex delta functions are defined by \cite{Witten:2004cp}
\be{delta-bar}
\bar\delta(z)=\delta(\mathrm{Re}\; z)\delta(\mathrm{Im}\; z)\d \bar z =\frac1{2\pi i}\dbar \frac 1z.
\ee
This leads to the formula:
\be{MHV-bar}
\cM^{\mathrm{grav}}_{\overline{\mathrm{MHV}}}(1,2,3)= 2\Lambda \frac{[1\, 2]^6}{[1\, 3]^2[2\, 3]^2} \left(2- \Lambda \Box_p\right) \delta^4\left(\sum p_i\right)
\ee
where $\Box_p$ is the wave operator in the momentum variable.\footnote{This latter comes about because of the term $\Lambda \varepsilon_{AB}\p h_1/\p\lambda_{A}\p h_2/\p \lambda_{B}$ in the Poisson bracket.  The derivative with respect to $\lambda_{A}$ when acting on $h_k$ as in \eqref{mom} can be re-expressed as derivatives with respect to $p_{A}$ which in turn lead to $\tilde{p}_{A'}\p/\p p_{AA'}$ when acting eventually on the momentum conserving delta function.

 Note that it does not matter which momentum variable is being differentiated as the delta-function is a function of the sum of all three.  The failure of translation invariance that we expect in de Sitter space arises essentially because of this $\Box_p$ term as it is the Fourier transform of a factor of $x^2$, and this rather more manifestly breaks translation invariance.  } This
reduces to the standard answer when divided by $\Lambda$ in the limit
as $\Lambda\rightarrow 0$.  This provides an explanation for the
observation of \cite{Brodel:2009ep} that the Einstein three point
MHV-bar amplitude sits inside the conformal supergravity three point
MHV-bar amplitude.

Berkovits and Witten \cite{Berkovits:2004jj} also identify an amplitude involving just three $V_f$s for conformal gravity, and after reduction to Einstein supergravity this gives 
$$
\cM(h_1,h_2,h_3)=\int_\PT I^{NI}I^{JK}I^{LM}\p^2_{IJ}h_{1}\; \p^2_{KL}h_2\; \p^2_{MN}h_3\,\wedge\Omega .
$$
In particular, this determines a coupling linear in the scalar and
quadratic in the ASD spin-2 part of gravity consistent with a $\phi
W^2$ term in the action.  This term is clearly absent for $\cN=0$ and
its appearance at $\cN=4$ is not consistent with Einstein
supergravity, and is therefore spurious. We also note that this is a
contribution that violates the relationship between the MHV degree and
the degree of the the map involved.  

All other possible insertions of vertex operators vanish at degree
zero as any occurrence of $\rd Z$ needs to be contracted with a $Y$ to
obtain a non-zero answer as $Z$ is constant at degree zero.

\subsection{Reduction to the SD twistor actions}

The degree zero sector couplings suggest SD twistor actions as follows.  Here the $f^I$ is regarded as a perturbation $\bar\p_f=\bar\p+f^I\p_I$ of the standard complex structure $\bar\p$ on $\PT$.  The $\bar Q$-closure requirement gives  $\dbar f^I=0$.  This combined with the degree-zero interactions gives the Berkovits-Witten self-dual conformal supergravity action \cite{Berkovits:2004jj}
$$
S^{\mathrm{C-SUGRA}}_{\mathrm{SD}}[f,g]=\int_{\PT} g_I\wedge (\dbar f^I + [f,\wedge f]^I)\wedge \Omega\, . 
$$
The equations of motion give the vanishing of $N^I(f):=\dbar f^I + [f,\wedge f]^I$, the Nijenhuis tensor associated to the almost complex structure $\bar\p_{f}=\bar\p+f^I\p_I$.   This vanishing implies the integrability $\dbar_f^2=0$ of the almost complex structure $\dbar_f$.  The other equation of motion is $\dbar_f g_I=0$.  
Via the supersymmetric nonlinear graviton construction, this leads to SD conformal supergravity with an anti-self-dual linear conformal supergravity field $W(x,\theta)$ corresponding to $g_I\d Z^I$ on the nonlinear SD background.  

When \eqref{Einstein-V} is substituted into the action, we obtain
contractions of the infinity twistor with itself in both the quadratic
term and the cubic term leading to $\Lambda Z^I\p_I h=2\Lambda h$ by
homogeneity in both terms (just as in the relation $X_{h}\lrcorner
\tau=2\Lambda h$ before).   These are off-shell relations and we
therefore obtain
$$
S_{\mathrm{SD}}^{\mathrm{SUGRA}}[h,\tilde{h}]=2\Lambda \int_{\PT} \tilde{h}\wedge (\dbar h + \{h,\wedge h\})\wedge \Omega\, , 
$$
and, up to the factor of $2\Lambda$,  this is the self-dual Einstein supergravity action as found in \cite{Mason:2007ct}.
Once divided by $\Lambda$, this action gives the SD sector of supergravities both when $\Lambda=0$ and $\Lambda\neq 0$.

\subsection{Amplitudes at degree 1}

For the non-supersymmetric case $V_h$ has no fermonic variables and $V_{\tilde h}$ has four.    Since there are now $8$ fermionic
integrations, we must have precisely  two $V_{\tilde h}$ insertions but can have
as many $ V_h$ insertions as we like, and this leads to the MHV
amplitudes.  For the case of the three point MHV amplitude 
\be{3pt-MHV-ts}
\cM^{\mathrm{C-SUGRA}}_{\mathrm{MHV}}(1,2,3)=\int_{\CM_{1,3}}\d\mu_1 \la V_h V_{\tilde h}V_{\tilde h}\ra\, , 
\ee
since the conformal supergravity twistor-string is thought to be
parity invariant, we expect to obtain the parity conjugate of the
MHV-bar amplitude \eqref{MHV-bar} (this must be the case if, as we have been assuming, the twistor string correctly reproduces conformal supergravity tree amplitudes).  This can be compared to the calculations of \cite{Dolan:2008gc}.  They do show that
the spin two part of the amplitude will vanish when $\Lambda=0$ (this
is the vanishing of their $\la e_2 e_{-2}e_{-2}\ra$ component) as we expect.
However, they do not check the nonvanishing components of the `$\int\la V_gV_gV_f\ra$' amplitude (this would in particular include their $\la e_2' e_{-2}'e_{-2}'\ra$ component).

At MHV, the contraction of $V_h$ with $\tau$ vanishes.  This can be seen as follows.  We first note that
\begin{equation*}
\la Z^{I}(\sigma) V_f\ra=\int_{\CP^1}\frac{\D\sigma'}{(\sigma\cdot\sigma')}\frac{(\xi\cdot\sigma)^2}{(\xi\cdot\sigma')^2}f^{I}(Z(\sigma')).
\end{equation*}
The contraction with $\tau$ is therefore
$$
\la \tau (\sigma) V_f\ra = I_{IJ} \D\sigma\int_{\CP^1}\frac{\D\sigma'}{(\sigma\cdot\sigma')^2}\frac{(\xi\cdot\sigma)}{(\xi\cdot\sigma')^2}\left(2\xi\cdot\sigma' \, Z^{I }(\sigma)- \xi\cdot\sigma \,Z^{I}(\sigma')\right)f^{J}(Z(\sigma'))
$$ 
so if $f^I=I^{IJ}\p_J h$ is an Einstein wave function, we obtain a contraction between two infinity twistors which gives $I_{IJ}I^{JK}=\Lambda \delta _I^K$ so that we obtain 
\begin{multline*}
\la \tau (\sigma) V_h\ra=\Lambda\D\sigma \int_{\CP^{1}}\frac{\D\sigma'\;(\xi\cdot\sigma)}{(\sigma\cdot\sigma')^{2}(\xi\cdot\sigma')^{2}}\left(2(\xi\cdot\sigma')Z^{I}(\sigma)-(\xi\cdot\sigma)Z^{I}(\sigma')\right)\partial_{I} h(\sigma') \\
=2\Lambda \D\sigma \int_{\CP^1}\frac{\D\sigma'\;(\xi\cdot\sigma)}{(\sigma\cdot\sigma')^{2}(\xi\cdot\sigma')^{2}}\left((\xi\cdot\sigma')\sigma\cdot\partial' h(\sigma')-(\xi\cdot\sigma)h(\sigma')\right) \\
=2\Lambda\D\sigma \sigma_{A}\int_{\CP^1}\frac{\partial}{\partial\sigma'_{A}}\left(\frac{\D\sigma'\;(\xi\cdot\sigma)h(\sigma')}{(\sigma\cdot\sigma')^{2}(\xi\cdot\sigma')}\right) \\
= 2\Lambda \D\sigma \sigma_{A}\int_{\CP^1}\partial' \left(\frac{\sigma^{\prime A}(\xi\cdot\sigma)h(\sigma')}{(\sigma\cdot\sigma')^{2}(\xi\cdot\sigma')}\right)=0.
\end{multline*}
In the second line we have used the homogeneity  relation $Z\cdot \p h=2h$ and the chain rule together with the linearity of $Z(\sigma')$ in $\sigma'$ to deduce that $\sigma\cdot \p_{\sigma'}h(Z(\sigma'))=Z^I(\sigma)\p_I h(Z(\sigma'))$.  
Thus contractions of $\tau$ with $V_h$ vanish\footnote{ This calculation can also be understood in terms of building a Picard-iterative solution to the equation
\begin{equation*}
\dbar Z^{I}(\sigma)=I^{IJ}\partial_{J}h(Z(\sigma))
\end{equation*}  for a rational curve of degree 1 with respect to the complex structure deformed by the Hamiltonian vector field of $h$.  The contraction of $V_h$ with $\tau$ gives the first order deformation of $\tau$ under the deformation and so its vanishing corresponds to the preservation of $\tau$ under this deformation.
Further details will appear in \cite{Adamo:2012}.}.  

 The only Wick contractions are therefore of the form:
\begin{equation*}
I^{IJ}\left\la Y_{2\;I}\tilde{h}_{3}\right\ra_{d=1}\partial_{2\;J}h_{2}=\frac{\D\sigma_{2}(\xi\cdot \sigma_{3})^{2}}{(\sigma_{2}\cdot\sigma_{3})(\xi\cdot\sigma_{2})^{2}}I^{IJ}\partial_{I}\tilde{h}_{3}\partial_{J}h_{2}.
\end{equation*}
Hence, \eqref{3pt-MHV-ts} can be written:
\be{3pt-MHV2}
\cM^{\mathrm{C-SUGRA}}_{\mathrm{MHV}}(1,2,3)=\int_{\CM_{1,3}}\d\mu_1\;(U^{2})^{2}\prod_{j}\D\sigma_{j}\frac{(\xi\cdot \sigma_{3})^{2}\tilde{h}_{1}}{(\sigma_{2}\cdot\sigma_{3})(\xi\cdot\sigma_{2})^{2}}I^{IJ}\partial_{I}h_{2}\partial_{J}\tilde{h}_{3}+(1\leftrightarrow 3).
\ee

Using $Z^{I}=U^{I A}\sigma_{A}$, we can set
\begin{equation*}
\partial_{J}\tilde{h}_{3}=\frac{\sigma^{B}_{1}}{(\sigma_{3}\cdot\sigma_{1})}\frac{\partial \tilde{h}_{3}}{\partial U^{I B}}.
\end{equation*}
Inserting this into \eqref{3pt-MHV2}, we can integrate by parts with respect to $U$.  Our choice means that $\frac{\partial}{\partial U}$ annihilates $\tilde{h}_{1}$ as well as $I^{IJ}\partial_{I}h_{2}$, since this vector is divergence-free.  Hence, the only contribution we obtain is when the derivative hits the $(U^2)^2$ factor leading to 
\begin{multline*}
4\int \d\mu_{1}\; U^{2} I_{IK}Z_{1}^{K} \frac{(\xi\cdot\sigma_{3})^{2}}{(\sigma_{3}\cdot\sigma_{1})(\sigma_{2}\cdot\sigma_{3})(\xi\cdot\sigma_{2})^{2}}I^{IJ}\tilde{h}_{3}\partial_{J}h_{2} \tilde{h}_{1}\prod_{j}\D\sigma_{j} +(1\leftrightarrow 3) \\
=-4\Lambda \int \d\mu_{1}\;U^{2}\frac{(\xi\cdot\sigma_{3})^{2}}{(\sigma_{3}\cdot\sigma_{1})(\sigma_{2}\cdot \sigma_{3})(\xi\cdot\sigma_{2})^{2}}\tilde{h}_{3}\;Z_{1}\cdot\partial_{2} h_{2}\tilde{h}_{1}\prod_{j}\D\sigma_{j}+(1\leftrightarrow 3).
\end{multline*}

When $Z^{I}(\sigma)$ is a degree one function in $\sigma$,
with $Z_i=Z(\sigma_i)$, the chain rule gives 
\begin{equation*}
Z_{1}\cdot\frac{\partial}{\partial Z_{2}}f(Z_2)= \sigma_{1}\cdot\frac{\partial}{\partial\sigma_{2}} f(Z(\sigma_2))\, .
\end{equation*}
This enables us to integrate by parts with respect to $\D\sigma_{2}$, obtaining
\begin{multline*}
4\Lambda\int \d\mu_{1}\;U^{2}\frac{(\xi\cdot\sigma_{3})^{2}}{(\sigma_{3}\cdot\sigma_{1})(\sigma_{2}\cdot\sigma_{3})(\xi\cdot\sigma_{2})^{2}}\left(\frac{(\sigma_{3}\cdot\sigma_{1})(\xi\cdot\sigma_{2})-2(\xi\cdot\sigma_{1})(\sigma_{2}\cdot\sigma_{3})}{(\sigma_{2}\cdot\sigma_{3})(\xi\cdot\sigma_{2})}\right) \\
\times h_{2}\tilde{h}_{1}\tilde{h}_{3}\prod_{j}\D\sigma_{j} +(1\leftrightarrow 3).
\end{multline*}
Two applications of the Schouten identity reduces this to
\be{3ptMHV-N=4}
\cM^{\mathrm{SUGRA}}_{\mathrm{MHV}}(1,2,3)=2\Lambda \int_{\CM_{1,3}}\d\mu_1  U^2 \frac {(\sigma_3\cdot \sigma_1)^2}{(\sigma_1\cdot\sigma_{2})^2(\sigma_2\cdot \sigma_3)^2 } \tilde h_1h_2 \tilde h_3 \D\sigma_1 \D\sigma_2 \D\sigma_3\, ,
\ee
and on insertion of moment eigenstates for the $h_i$,  this evaluates directly to give the parity conjugate of \eqref{MHV-bar}.  This follows by observing that $U^2=(2-\Lambda x^2)$ so that once the $d^4x$ is integrated out to obtain the momentum delta-functions, the $x^2$ factor becomes a $\Box_p$ on the momentum delta function. 

\subsubsection*{\textit{Spurious amplitudes}}

At degree 1, we can
also construct amplitudes with say $n$ $V_{\tilde h}$ insertions.  Here there are no contractions and each $V_{\tilde h}$ integral is a Penrose transform evaluation that directly  gives $\tilde{\Phi}(x,\theta)$ on the degree-1 line corresponding to $(x,\theta)$.  Thus the amplitude is simply $\int \rd^{4|8}x \tilde \Phi(X,\theta)^n$.  This contains terms such as $\int \rd^4x \tilde \phi^{n-2}\tilde W^2$ (the $n=3$ case being the parity conjugate of the degree zero $\la V_hV_hV_h\ra$ amplitude).  
These sum to give
$$
\int\d^{4|8}x \e^{\tilde \Phi(x,\theta)}=\int\d^4x \e^{\tilde \phi}(\tilde W^2)+\ldots
$$
This is part of the expected  action for conformal supergravity and so is not consistent with amplitudes from an Einstein supergravity action.

Amplitudes with just $V_h$ insertions are studied in \cite{Ahn:2005es}. These can come in at different degrees, and when
local on space-time, will be the parity conjugate of those arising from
multiple  $V_{\tilde h}$ insertions.  These also correspond to conformal supergravity
interactions that do not occur in Einstein supergravity. 

These two classes of degree one non-MHV amplitudes are in a different conformal
supersymmetry representation than the single nontrivial MHV Einstein gravity amplitude.  Indeed in conformal supergravity there are nontrivial amplitudes of any MHV degree at degree one.  However, only the MHV amplitude contains a nontrivial $\cN=0$ Einstein gravity amplitude and its extension to $\cN=4$ supersymmetry will therefore be the nontrivial Einstein supergravity MHV amplitude.  None of the N$^k$MHV conformal supergravity amplitudes for $k\neq 0$ will play a part in  Einstein supergravity at degree 1 and must be ignored.   

\subsection{Higher degree}

We have made the assumption that the twistor-string
formulae do correctly give conformal supergravity at tree level as
they are believed to. Thus with the vertex operators \eqref{Einstein-V}, \eqref{0susy} appropriate to
Einstein gravity without supersymmetry, we will obtain gravity amplitudes by the Maldacena argument.  By counting fermionic integrals against the fermionic coordinates appearing in the vertex operators, it follows that these will
only be nontrivial when the degree is $d=1+k$, with $k$ the MHV degree.  The
$\cN=4$ completion of these N$^k$MHV amplitudes is then precisely what
we obtain 
 from the ansatze
\be{N=4}
\cM_k^{\cN=4}(1,\ldots,n)=\frac1\Lambda \int_{\CM_{k+1,n}}\d\mu_{k+1}\la V_{\tilde{h}_1}\ldots V_{\tilde{h}_{k+2}}V_{h_{k+3}}\ldots V_{h_n}\ra_{d=k+1}\, .
\ee
These amplitudes are in the same irreducible representation of $\cN=4$ SUSY as the standard nontrivial $\cN=0$ Einstein amplitudes, and therefore must be the nontrivial Einstein $\cN=4$ SUGRA amplitudes.  The other possible amplitudes that one might inherit from conformal gravity, in which the MHV-degree of the amplitude is not related to the degree of the rational curve, $d\neq k+1$, will be spurious like those discussed above at degree zero and degree one.
We therefore conclude that \eqref{N=4} gives the tree-level S-matrix of Einstein gravity with $\Lambda\neq 0$.   

However, by construction, each vertex operator is linear in $\Lambda$, so an $n$ particle  amplitude is a  polynomial in $\Lambda$ of degree $n$ with vanishing $\Lambda^0$ part by the Maldacena argument.  Thus \eqref{N=4} also makes sense at $\Lambda=0$ and must by continuity give the correct answer there. It has so far proved difficult to make the expected overall factor of $\Lambda$ in these amplitudes explicit except at degrees 0 and 1.  


\section{$\cN=8$ Supergravity and Recursion}
\label{Recursion}

For Einstein supergravity we expect to be able to obtain formulae all the way up to $\cN=8$.  The twistor-string approach does not immediately give any guidance on this and in the following we obtain formulae based on supersymmetric BCFW recursion for gravity \cite{ArkaniHamed:2008gz} and its translation into twistor space \cite{Mason:2009sa}. Here we rewrite those formulae in a notation that is more suggestive of twistor-string theory (and without the restriction to split signature) and extend them to $\Lambda\neq 0$.\footnote{For this extension to be valid, one must also prove the extension of the result of \cite{ArkaniHamed:2008yf} that the BCFW shift of Einstein amplitudes with cosmological constant falls off appropriately as the parameter goes to infinity.  We outline how this can be done in appendix \ref{BCFWapp}.}  We will be working directly with the Einstein amplitudes so the overall factor of $\Lambda$ present in the formulae above will be absent and we will be able to take $\Lambda\rightarrow 0$ directly.  However, we will only be able to obtain twistor-string-like forulae at degree zero and one.

Hereon, we will work in $\cN=8$ supertwistor space $\T_{[8]}$ so that $Z^I=(Z^\alpha,\chi^a)$ where now $a=1,\ldots ,8$ and the corresponding holomorphic volume form $\Omega_{[8]}=\D^{3|8}Z$ now has weight $-4$ (so $\PT_{[8]}$ is no longer Calabi-Yau). We can embed the above $\cN=4$ superfields into the $\cN=8$ framework by setting
\be{4in8}
H=h+\chi^5\chi^6\chi^7\chi^8 \tilde h\, .
\ee
A generic $H$ of homogeneity degree two will encompass the full $\cN=8$ linear gravity supermultiplet: via the Penrose transform, the different coefficients of the powers of $\chi$ will correspond to the different component fields of the multiplet.

The BCFW recursion framework in \cite{Mason:2009sa} was based on a split signature framework in which the twistors are totally real and in which wavefunctions $H_i$ were not left arbitrary, but were supported at a single twistor.  The $n$-point amplitude was thereby represented as a function (more accurately a distribution) on $n$ copies of twistor space.  It is easily translated into the complex framework used here by working on complex twistor spaces, and replacing the real delta functions $\delta^{3|8}(Z_i,Z)$ by complex ones 
\be{delta-fn-w}
\bar\delta^{3|8}(Z_i,Z)=\int_\C \frac{\d s}{s^3}\bar\delta^{4|8}( Z_i+sZ)\, , \quad \mbox{ where } \quad \bar\delta^{4|8}(Z):=\prod_{\alpha=0}^3 \bar\delta^1(Z^\alpha)\prod_{a=1}^8 \chi^a
\ee
with the delta functions $\bar\delta^1$ given by \eqref{delta-bar} (and we use the fermionic relation $\delta^{0|1}(\chi)=\chi$).  Thus, we take $H_i=\bar\delta^{3|8}(Z_i,Z)$ which is understood as a $(0,1)$-form in the $Z$-variable and a $(0,2)$-form in the $Z_i$ variable so that it can be integrated against ordinary wave functions in the $Z_i$ variable so as to give back the standard formulae above.

The recursion is seeded by the three point MHV and MHV-bar amplitudes.  We have the formula \eqref{MHV-bar-T} for the MHV-bar amplitude with $\cN=4$ supersymmetry; removing the factor of $\Lambda$ and extending to $\cN=8$ gives
\be{MHV-bar-N=8}
\cM^{\cN =8}_{\overline{\mathrm{MHV}}}(1,2,3)= \int_{\PT_{[8]}}  H_3 \{{H}_1,{H}_2\} \wedge \Omega_{[8]} \, .
\ee
For the MHV amplitude from \eqref{3ptMHV-N=4} we similarly obtain
\be{3ptMHV}
\cM^{\cN =8}_{\mathrm{MHV}}(1,2,3)=\int_{\CM_{1,n}}\hskip-0.3cm\d\mu_1  \frac {H_1\tau_1 \; H_2 \D\sigma_2 \; H_3 \D\sigma_3}{(\sigma_1\cdot\sigma_{2})^2(\sigma_2\cdot \sigma_3)^2 (\sigma_3\cdot \sigma_1)^2} =\int_{\CM_{1,n}}\hskip-0.3cm\d\mu_1 \la U_0U_1\ra \prod_{i=1}^3 \frac { H_i \D\sigma_i }{(\sigma_i\cdot\sigma_{i+1})^2}  
\ee
where we use the notation 
$$
Z(\sigma)=U_0\sigma_0+U_1\sigma_1\, , \qquad \la U_0U_1\ra=
I_{IJ} U_0^IU_1^J\, .
$$  
Equivalent formulae are also obtained in the split signature context in \cite{Mason:2009sa} by half-Fourier transform (or alternatively in this complex context, we can substitute in momentum eigenstates \eqref{mom} to obtain the standard momentum space formulae).

Higher point amplitudes can be obtained by BCFW recursion.  The BCFW momentum shift on twistor space yields
$$
\cM(Z_1,\ldots,Z_n)\rightarrow \cM(Z_1,\ldots,Z_n+tZ_1)\, .
$$
On momentum space a simple residue argument  leads to  the BCFW recursion relation \cite{Britto:2004ap,Britto:2005fq} and this was extended to gravity in \cite{Bedford:2005yy, Cachazo:2005ca, Benincasa:2007qj} incorporating supersymmetry in \cite{ArkaniHamed:2008gz}.   When reformulated on twistor space this becomes
\be{recursion}
\cM(Z_1,\ldots,Z_n)=\sum_{L
,R}\int_{\C^*\times \PT_{[8]}}\hskip-.5cm \D^{3|8}Z \frac{\d t}t  \cM_L(Z_1, Z_2,\ldots,Z_i,Z)\cM_R(Z,Z_{i+1},\ldots,Z_n+tZ_1)
\ee
where the sum is over all $1<i<n-1$ and permutations fixing $1$ and $n$.  

In the solution to the recursion relations, a particularly important
role is played by the contributions in which either $\cM_L$ or $\cM_R$
is a three point amplitude.   Up to various shifts, these are the main
terms involved in solving the recursion relations inductively.    The
special three-particle kinematics implies that these
contributions are only nontrivial when $\cM_L$ is MHV or $\cM_R$ is
the MHV-bar.  In  these cases, the  integration were performed
explicitly in section 6 of \cite{Mason:2009sa}.   To generate the $n$-point MHV
amplitude we just need the recursion following from taking $\cM_R$ to
be the three point 
MHV-bar amplitude leading to the recursion
\begin{multline}\label{hgs-term}
\cM(Z_1,\ldots,Z_n)= I^{IJ}\p_{nI}\bar\delta^{2|8}(Z_1,Z_{n-1},Z_n)
\p_{n-1\, J}\cM(1,\ldots,n-1)\\ + \mbox{Perms} \, \{2,\ldots,n-1\}
\end{multline}
where 
$$
 \bar\delta^{2|8}(Z_1,Z_{n-1},Z_n)=\int_{\C\times\C}\frac{\d s\d t}{s^2t} \bar\delta^{4|8}(Z_n+ s Z_{n-1}+tZ_1)
$$ 
and $\bar\delta^{2|8}$ has homogeneity $(0,1,3)$ in its respective arguments.  The $\bar\delta^{2|8}$ has the effect of ensuring that $Z_n$ is collinear with $Z_1$ and $Z_{n-1}$ and so in the MHV case, the amplitude is supported where the twistors $Z_i$ are all collinear.  Rewriting the $(s,t)$ in terms of $\sigma_n$ by $\sigma_n=s\sigma_{n-1}+t\sigma_1$, we can reexpress the resulting formula for the full MHV amplitude in a twistor-string format as
\begin{eqnarray}
\cM^{\cN =8}_{0}(Z_1,\ldots,Z_n)&=&\int_{\CM_{1,n}}
\d\mu_1  \left(\prod_{i=4}^n 
\frac{
[\p_i\, \p_{i-1}] H_i \D\sigma_i }{\sigma_i\cdot\sigma_{i-1}}\right)\frac{H_3 \D\sigma_3 \; H_2 \D\sigma_2 \; H_1\tau_1}{(\sigma_3\cdot\sigma_2)^2(\sigma_2\cdot\sigma_1)^2(\sigma_1\cdot\sigma_n)^2}
\nonumber \\
&& +\mbox{Permutations of 2 to $n-1$} \label{grav-MHV-BCFW}
\end{eqnarray}
where $H_i=\bar\delta^{3|8}(Z_i,Z(\sigma_i))$, $[\p_n\, ,
\p_{n-1}]=I^{IJ}\frac{\p }{\p Z_n^I}\frac{\p }{\p Z_{n-1}^J}$ and the
terms in the product are ordered with increasing $i$ to the left.  (In
order to be fully comparable to earlier formulae, we must multiply by
generic wave-functions and integrate out the $Z_i$ and we also need to
use the arguments of \cite{Drummond:2009ge} to use a cyclically ordered version
of the recursion in which we take just the one term in \eqref{hgs-term} and then sum the final result over all permutations of $2$ to $n-1$.)

It is instructive to compare this to \eqref{N=4} at $k=0$.  In that case we obtain
\begin{eqnarray} \nonumber
\cM^{\cN =4}_{0}(1,\ldots,n)&=&\int_{\CM_{1,n}}\d\mu_{1}\la V_{h_n} \ldots V_{h_4} V_{h_3} \tilde h_{2}\tau_2 \tilde h_{1}\tau_1\ra_{d=1} +\\ &&+\mbox{permutations of  3 to n and $1\leftrightarrow 2$}\, ,\label{E-sugra0}
\end{eqnarray}
where $V_{h_3}=Y_{3I}I^{IJ}\p_Jh_3$, etc.  This formula, arising as it does from conformal supergravity, will have an additional factor of $\Lambda$.  It also has two extra derivatives (each $V_h$ essentially involves two derivatives).   We can see that  when for example $Y_3$ is contracted with the $\d Z_3$ contained in $\tau_3$, a $\Lambda$ will appear and two derivatives will be lost via a process such as \eqref{reduction} so we will obtain a term such as that displayed in \eqref{grav-MHV-BCFW}.  Equations \eqref{E-sugra0} and \eqref{grav-MHV-BCFW} are proved to be equivalent up to a factor of $\Lambda$ in \cite{Adamo:2012}.

\section{Conclusion and discussion}
We have seen that using the twistor-string tree formulae \eqref{N=4} for Einstein supergravity with $\cN=0$ and $4$, tree amplitudes exist for all degrees $d$ of the rational curve and MHV degrees $k$ with the expected relationship $d=k+1$ between the degree of the rational curve and the MHV degree.  By construction these are polynomials of degree $n$ in $\Lambda$ that, by the Maldacena argument vanish when $\Lambda=0$, so they can be divided by $\Lambda$ to yield answers that are correct also in the limit as $\Lambda\rightarrow 0$.  This provides much more evidence than we have hitherto had that a twistor-string theory for Einstein gravity can be made to work.  However, many challenges remain to a proper understanding of the structures even at tree level.

There remain issues with the computation of the conformal field theory correlators on the worldsheet such as $\la V_{\tilde h_1} \ldots V_{\tilde h_{k+2}}V_{h_{k+3}} \ldots V_{h_n}\ra$.   Equation \eqref{3ptMHV-N=4} was obtained from the twistor action for conformal gravity rather than via a direct CFT calculation and is hard to show from the naive rules given earlier (although it does follow from a more geometric path-integral argument following the strategy outlined in \cite{Nair:2007md}, although it doesn't arise directly from the formula given there).  These world-sheet conformal field theory calculations need a more systematic understanding in order to make this a useful tool for the study of  Einstein amplitudes.

By the Maldacena argument, the formulae from conformal supergravity must vanish in the $\Lambda\rightarrow 0$ limit.
This vanishing as $\Lambda\rightarrow 0$ is not manifest in the above formula.  We can choose a non-degenerate infinity twistor (corresponding to a gauged supergravity) with $I_{IJ}I^{JK}=\Lambda\delta^K_I$ so that any contraction between $I^{IJ}Y_J$ and a $\tau$ will yield a factor of $\Lambda$.  This vanishing as $\Lambda\rightarrow 0$ will follow if we can show that the terms with no contractions between a $Y$ and a $\tau$ combine so as to vanish.  We also expect the number of derivatives in the formula to drop by two in the process as the degree zero and one examples illustrate, and as suggested by the formulae from BCFW recursion at higher MHV degree.

At a more basic level, we have worked under the assumption that the original twistor-string theories of Witten and Berkovits correctly compute conformal supergravity amplitudes at tree level.  Despite the fact that the gauge theory formula \eqref{gauge-tree} has been around since 2004, it has only recently been proved \cite{Skinner:2010cz,Dolan:2011za} and, as remarked above, the technicalities for conformal gravity are harder.

Another challenge is to extend these formulae and indeed the twistor-string theory to $\cN=8$.  
Indeed we can naively extend the twistor-string action to $\cN=8$ (with the fermionic components of both $Y$ and $Z$ extended to $\cN=8$). 
We can then consider the naive extensions of our two types of vertex operator
\be{N=8-V}
V_H=Y_II^{IJ}\p_JH\, , \qquad \widetilde V_H=H \wedge\tau \, .
\ee
However, whereas $V_H$ has the expected homogeneity degree zero, $\widetilde V_H$ now has weight 4 in $Z$.   So although $V_H$ can be added to the action as before, this is no longer true of $\widetilde V_H$ which will no longer be a straightforward $(1,1)$-form on the worldsheet beyond degree $0$ and cannot be integrated for $d>0$.  Nevertheless, despite this crude mismatch, these weights are good for enforcing the relationship $d=k+1$ required to balance the weight of the instanton moduli measure against those of $\widetilde V_H$ as the instanton moduli space is no longer Calabi-Yau.

In a subsequent paper \cite{Adamo:2012} we will follow an approach based again on the Maldacena argument, but using the twistor action for conformal supergravity presented in \cite{Mason:2005zm} and extending the ideas from \cite{Mason:2008jy} to gravity with a cosmological constant.  Here the Berkovits-Witten self-dual action is supplemented by a term that generates the MHV amplitudes and provides a twistor action for full conformal supergravity. This has the interpretation as a string-field theory action for twistor-string theory but relies only on the degree 0 and 1 contributions.  In \cite{Adamo:2012}, we will see that a similar phenomenon occurs, and when evaluated on Einstein data, it gives $\Lambda$ times the appropriate term for Einstein gravity essentially via a straightforward integration by parts argument.  This then extends the construction of the Einstein MHV amplitude in \cite{Mason:2008jy} to non-zero cosmological constant and gives a more robust derivation of that action.

BCFW recursion can also be used to obtain formulae at higher MHV degree, indeed all the anti-MHV amplitudes were obtained in \cite{Mason:2009sa}. However, these will be supported on configurations  of $2k+1$ intersecting lines at N$^k$MHV \cite{Mason:2009sa,Korchemsky:2009jv, Bullimore:2009cb} and so do not directly lead to twistor-string like formulae, although relations can in principle be obtained using the ideas of \cite{Skinner:2010cz} and \cite{Bullimore:2009cb}.

\subsubsection*{\textit{Acknowledgments}}

TA is supported by a National Science Foundation (USA) Graduate Research Fellowship and by Balliol College; LM is supported by a Leverhulme Fellowship. 


\appendix

\section{BCFW with Cosmological Constant}
\label{BCFWapp}

In this paper (in particular, in section \ref{Recursion}), we have assumed that BCFW recursion holds for gravitational amplitudes on a background with cosmological constant, $\Lambda$.  As first illustrated in \cite{Britto:2005fq}, BCFW recursion can be derived by picking two external momenta for a scattering amplitude and analytically continuing them with a complex variable $z$ while keeping them on-shell and maintaining overall momentum conservation.  The amplitude then becomes a complex function $\cM(z)$: it has simple poles wherever internal propagators go on-shell, and $\cM(0)$ is the original amplitude.  These simple poles correspond to the terms arising in the BCFW recursion, so provided $\cM(z\rightarrow\infty)$ vanishes, Cauchy's theorem implies that
\begin{equation*}
0=\frac{1}{2\pi i}\int \frac{\d z}{z}\cM(z)=\cM(0)+\mbox{BCFW terms},
\end{equation*}
as desired.       

In the $\Lambda=0$ case, it was proven that $\cM(z\rightarrow\infty)=0$ using an elegant background field method in \cite{ArkaniHamed:2008yf}.  With $\Lambda\neq 0$, $\cM(z)$ still has simple poles corresponding to propagators going on-shell, so the only potential subtlety arises with the fall-off as $z\rightarrow\infty$, and it suffices to show that the methods of \cite{ArkaniHamed:2008yf} still work.  In the large $z$ regime, we are interested in quadratic fluctuations on a classical background, where the fluctuations correspond to the two shifted particles and the soft background looks like de Sitter space.  For our gravitational amplitudes, this entails inserting a metric $g_{\mu\nu}+h_{\mu\nu}$, and extracting the portion which is quadratic in $h$ \cite{Christensen:1979iy}:
\begin{multline*}
\cL_{\mathrm{quad}}=\sqrt{-g}\left[\frac{1}{4}\tilde{h}^{\mu\nu}(2R_{\mu\rho}g_{\mu\sigma}-2R_{\mu\rho\nu\sigma}-g_{\mu\rho}g_{\nu\sigma}\Box)h^{\rho\sigma}-\frac{1}{2}\nabla^{\rho}\tilde{h}_{\rho\mu}\nabla^{\sigma}\tilde{h}^{\mu}_{\sigma}\right. \\
\left. -\tilde{h}(R_{\rho\sigma}-\frac{1}{4}g_{\rho\sigma}R)h^{\sigma}_{\mu}-\frac{1}{2}\Lambda\tilde{h}^{\mu\nu}h_{\mu\nu}\right],
\end{multline*}  
where $\tilde{h}_{\mu\nu}=h_{\mu\nu}-\frac{1}{2}g_{\mu\nu}h$, and $h=g_{\mu\nu}h^{\mu\nu}$.  To this, we add the de Donder gauge-fixing term, as well as a Lagrangian density for a conformally-invariant scalar field, leaving us with:
\begin{multline*}
\cL_{\mathrm{quad}}=\sqrt{-g}\left[\frac{1}{4}\tilde{h}^{\mu\nu}(2R_{\mu\rho}g_{\mu\sigma}-2R_{\mu\rho\nu\sigma}-g_{\mu\rho}g_{\nu\sigma}\Box)h^{\rho\sigma}-\tilde{h}(R_{\rho\sigma}-\frac{1}{4}g_{\rho\sigma}R)h^{\sigma}_{\mu}\right. \\
\left. -\frac{1}{2}\Lambda\tilde{h}^{\mu\nu}h_{\mu\nu}+\frac{1}{2}g^{\mu\nu}\nabla_{\mu}\phi\nabla_{\nu}\phi -\Lambda\phi^{2}\right].
\end{multline*}

Now, we take our background metric $g_{\mu\nu}$ to be de Sitter, and implement the field re-definition used in \cite{Bern:1999ji}:
\begin{equation*}
h_{\mu\nu}\rightarrow h_{\mu\nu}+g_{\mu\nu}\phi, \qquad \phi\rightarrow \frac{h}{2}+\phi.
\end{equation*}
A bit of tensor algebra reveals that the quadratic Lagrangian transforms to become:
\begin{equation*}
\cL_{\mathrm{quad}}\rightarrow\sqrt{-g}\left[\frac{1}{4}g^{\mu\nu}\nabla_{\mu}h^{\sigma}_{\rho}\nabla_{\nu}h^{\rho}_{\sigma}-\frac{1}{2}h_{\mu\nu}h_{\rho\sigma}R^{\mu\rho\nu\sigma}+\frac{1}{2}g^{\mu\nu}\nabla_{\mu}\phi\nabla_{\nu}\phi -\Lambda\phi^{2}\right].
\end{equation*}
This transformation successfully eliminates all the trace terms, and after decoupling the re-defined scalar field, the Lagrangian is exactly the same as the one used in the flat background calculation.  From this point, the proof that $\cM(z\rightarrow\infty)$ vanishes follows in exactly the same fashion as in the $\Lambda=0$ case of \cite{ArkaniHamed:2008yf}.

\bibliographystyle{JHEP}
\bibliography{deSitter}

\end{document}